\documentclass[epj]{webofc}
\usepackage[utf8]{inputenc}
\usepackage[varg]{txfonts}   
\usepackage{booktabs}
\usepackage{xcolor}
\usepackage{braket}
\definecolor{darkred}{rgb}{0.4,0.0,0.0}
\definecolor{darkgreen}{rgb}{0.0,0.4,0.0}
\definecolor{darkblue}{rgb}{0.0,0.0,0.4}
\usepackage[bookmarks,linktocpage,colorlinks,
    linkcolor = darkred,
    urlcolor  = darkblue,
    citecolor = darkgreen]{hyperref}
\usepackage[font=small,format=plain,labelfont=bf,up,textfont=normal,up,justification=justified,singlelinecheck=true]{caption}
\usepackage{subfigure}
\wocname{EPJ Web of Conferences}
\woctitle{Lattice2017}
\begin{document}
%
\selectlanguage{english}
\title{%
Charm quark effects on the strong coupling extracted from the static force\footnote{Talk given at the 35th International Symposium on Lattice Field Theory, 18-24 June 2017, Granada, Spain. The title of the talk at the conference was ``Computation of $\alpha_{qq}$ in QCD ($N_f=2$) using Lattice Perturbation Theory''.}}
\author{
\firstname{Salvatore} \lastname{Cal\`i}\inst{1,2}\fnsep\thanks{Speaker, \email{scali@uni-wuppertal.de}} \and
\firstname{Francesco} \lastname{Knechtli}\inst{2} \and
\firstname{Tomasz}  \lastname{Korzec}\inst{2} \and 
\firstname{Haralambos}  \lastname{Panagopoulos}\inst{1} 
}
\institute{%
University of Cyprus, P.O. Box 20537, 1678 Nicosia, Cyprus 
\and
University of Wuppertal, Gau{\ss}str. 20, 42119 Wuppertal, Germany
}
\abstract{We compute the fermionic contribution to the strong coupling $\alpha_{qq}$ extracted from the static force in Lattice QCD up to order $g^4$ in perturbation theory. This allows us to subtract the leading fermionic lattice artifacts from recent determinations of $\alpha_{qq}$ produced in simulations of two dynamical charm quarks.

Moreover, by using a suitable parametrization of the $\beta_{qq}$-function, we can evaluate the charm loop effects on $\alpha_{qq}$ in the continuum limit.}
\maketitle
\section{Introduction}\label{intro} 

Many simulations of QCD are carried out with $2+1$ dynamical quarks, taking into account only the effects of light sea quarks (up, down, strange). So far, this kind of approach has provided really good results and its simulation costs are affordable with modern computer facilities.

However, the discovery of new charm-states in experiments like Belle, CLEO and BABAR has made charm physics really appealing in the last few years and including a dynamical charm quark in Lattice QCD simulations could help us understand better the properties of these new states.

Although nowadays new supercomputers allow to simulate the $2+1+1$ flavor theory, we know \cite{Bruno:2014ufa, Knechtli:2015lux} that the effects of a dynamical charm quark are small on low energy quantities, thus a high statistical precision is needed to disentangle them. Moreover, to resolve the small correlation length associated with a charm quark, really fine lattices are required to control the extrapolation to zero lattice spacing. Therefore it is interesting to understand for which kind of observables it is more worthwhile including a charm quark in simulations of QCD. 

In this work, we evaluate the charm loop effects on the strong coupling $\alpha_{qq}$, whose definition is
\begin{equation}
\alpha_{qq}(r) \equiv \frac{1}{C_F}r^2 V'(r),
\label{eq-aqq}
\end{equation}
where $C_F = 4/3$, $r$ is the distance between a static quark-antiquark pair and $V'(r)\equiv\frac{dV}{dr}$ is the derivative of the static potential $V(r)$ with respect to the distance. The static force $F(r)=V'(r)$ can also be used to measure a hadronic scale $r_0$ defined through $r^2F(r)\vert_{r=r_0}=1.65$ \cite{Sommer:1993ce}. Our first results about these studies can be found in \cite{Korzec:2016eko}, where we saw that charm loop effects on $\alpha_{qq}$ become significant at about 2 GeV. Here we extend our previous results, considering an additional dynamical ensemble, subtracting the leading fermionic lattice artifacts from our non-perturbative data and taking the continuum limit of $\alpha_{qq}$ through a convenient parametrization of the $\beta_{qq}$-function.

\section{Numerical setup}\label{sec-1}
To evaluate the dynamical charm effects on $\alpha_{qq}$, we compare QCD ($N_f=2$), with two heavy degenerate quarks having the charm mass $M_c$, to quenched QCD, namely QCD ($N_f=0$). 

As a lattice discretization, we consider Wilson's plaquette gauge action \cite{Wilson:1974sk} and a clover improved doublet of twisted mass Wilson fermions \cite{Sheikholeslami:1985ij,Frezzotti:2000nk}.
At maximal twist, the inclusion of a clover term is not needed for $\mathcal{O}(a)$ improvement of physical observables, but it reduces the $\mathcal{O}(a^2)$ lattice artifacts \cite{Dimopoulos:2009es}. Moreover, we choose open boundary conditions in the time direction and periodic boundary conditions in spatial directions.

To generate the dynamical ensembles, the bare coupling $g$ is chosen such that the lattice spacings cover the range $0.23\mbox{ fm} < a < 0.36\mbox{ fm}$, whereas the hopping parameter $\kappa$ is set to its critical value to achieve maximal twist. Finally, the twisted mass parameter $\mu$ is chosen such that the RGI mass in our simulations corresponds to the charm mass $M_c$.  The pure gauge theory is simulated at similar or smaller lattice spacings, setting the scale through the hadronic quantity $r_0$ as described in \cite{Necco:2001xg}. 

In particular, we study\footnote{Both quenched and dynamical ensembles have been produced using the program openQCD \cite{Luscher:2012av}, available at {\url{http://luscher.web.cern.ch/luscher/openQCD/}}.} three quenched ensembles at $\beta=6/g^2=6.34,6.672,6.90$ and three dynamical ensembles at $\beta=5.70,5.88,6.00$ and $M/\Lambda=4.87$, where $M/\Lambda$ is the ratio of the RGI mass to the $\Lambda$ parameter. All the simulation parameters of these ensembles are listed in Table 1 of Ref.~\cite{Knechtli:2017xgy}. For further details we refer to \cite{Korzec:2016eko, Knechtli:2017xgy}.

\section{Computation of the static force}\label{sec-force}
In this section we summarize the main steps that lead us to the computation of the strong coupling $\alpha_{qq}$ in QCD ($N_f=2$) at $M=M_c$ and QCD ($N_f=0$).

\subsection{Non-perturbative calculation}
From the definition of $\alpha_{qq}$ in the continuum, given in Eq. (\ref{eq-aqq}), it is clear that a lattice regularization of the derivative of the static potential $V(r)$ is needed. The most natural choice would be
\begin{equation}
F (r_{naive}) = \frac{1}{a} \left( V(Ra) - V(Ra - a) \right),\qquad \frac{r_{naive}}{a} = R - \frac{1}{2}, 
\label{eq-fnaive}
\end{equation}  
where the static potential $V(Ra)$ can be extracted from the expectation value of a rectangular Wilson loop $W(R,T)$ in the limit of infinite time separation, $T\rightarrow\infty$.
However, it has been shown \cite{Sommer:1993ce,Necco:2001xg} that it is better to introduce an improved distance $r_I$ such that the static force has no cutoff effects at the tree-level in perturbation theory, namely
\begin{equation}
F (r_{I}) = \frac{1}{a} \left( V(Ra) - V(Ra - a) \right) = C_F\frac{g^2}{4\pi r^2_I} + \mathcal{O}(g^4a^2).
\label{eq-flat}
\end{equation}
A table of the improved distances for unsmeared links is provided in \cite{Necco:2001xg}.

Here, in order to reduce the typical gauge noise which affects the measurements of $V(Ra)$ at large distances we follow Ref.~\cite{Donnellan:2010mx}. In particular, before we measure the Wilson loops, all the gauge links are replaced by HYP2-smeared ones \cite{Hasenfratz:2001hp}, which correspond to the following choice of the smearing HYP-parameters: $\alpha_1 = 1.0$, $\alpha_2 = 1.0$, $\alpha_3 = 0.5$. The static potential $aV(Ra)$ is extracted with great accuracy by smearing the initial and final lines of rectangular Wilson Loops up to four levels of HYP-smearing and then solving a generalized eigenvalue problem \cite{Blossier:2009kd}. Since the improved distances $r_I$ depend on the static quark action, we make use of the values of $r_I$ for HYP2-smeared links listed in Table 2 of \cite{Donnellan:2010mx}. Wilson loops have been computed using B. Leder's program available at \url{https://github.com/bjoern-leder/wloop/}.

\subsection{Perturbative calculation}
One of the goals of this work is to subtract the leading fermionic lattice artifacts from the non-perturbative measurements of $\alpha_{qq}$ realized on our dynamical ensembles.

For this purpose, first we compute the fermionic contribution to the static force up to order $g^4$ in perturbation theory. Then we extract the leading lattice artifacts comparing our calculations to the one-loop predictions of the continuum theory \cite{Melles:2000dq}. Such a computation has been performed in \cite{Athenodorou:2005hi,Athenodorou:2011zp} for clover improved Wilson fermions using unsmeared links. Here we adopt the same strategy, but we have to consider a clover improved doublet of twisted mass Wilson fermions and HYP2-smeared links. We just summarize the main ideas and we refer to \cite{Athenodorou:2005hi,Athenodorou:2011zp} for further details.

The perturbative expansion of a Wilson loop $W(R,T)$ can be written as
\begin{equation}
W(R,T) = 1 - g^2W_2(R,T) - g^4W_4(R,T) + \mathcal{O}(g^6).
\end{equation}
$W_2(R,T)$ involves only gluons, whilst $W_4(R,T)$ can be splitted into two terms
\begin{equation}
W_4(R,T)=W_4^g(R,T) + W_4^f(R,T), 
\end{equation}
where $W_4^g(R,T)$ comes from the pure-gauge theory and $W_4^f(R,T)$ is a purely fermionic contribution.

Once $W_2(R,T)$ and $W_4(R,T)$ have been calculated, it is possible to access the perturbative expansion of the static potential and the static force up to order $g^4$. Finally, converting to the $\overline{\mbox{MS}}$ scheme we can rewrite $F(r_I)$ as
\begin{equation}
F(r_I) = \frac{C_F\alpha_{\overline{MS}}(1/r_I)}{r_I^2}\left[ 1 + f_1(z,a/r_I)\alpha_{\overline{MS}}(1/r_I) + \mathcal{O}(\alpha^2_{\overline{MS}})\right],
\end{equation}
with
\begin{equation}
f_1(z,a/r_I) = f_{1,g}(a/r_I) + \sum_{i=1}^{N_f}f_{1,f}(z_i,a/r_{I}),\quad   z_i = z = r_Im_i,
\end{equation}
where $N_f$ is the number of flavors and $m_i$ are the quark masses. Since the corresponding continuum expressions $f_{1,g}(0)$ and $f_{1,f}(z,0)$ are known \cite{Melles:2000dq,Athenodorou:2011zp}, it follows that the relative lattice artifacts can be written as 
\begin{equation}
\frac{F(r_I) - F_{cont}(r_I)}{F_{cont}(r_I)} = \left(\delta^{(1,g)}_F(a/r_I) + \sum_{i=1}^{N_f}\delta^{(1,f)}_F(z_i,a/r_I)\right)g^2_{\overline{MS}}(1/r_{I}) + \mathcal{O}(g^4_{\overline{MS}}),
\label{eq-rel-lat-art}
\end{equation}
where
\begin{equation}
4\pi\delta^{(1,g)}_F(a/r_I) = f_{1,g}(a/r_I) - f_{1,g}(0),\quad 4\pi\delta^{(1,f)}_F(z_i,a/r_I) = f_{1,f}(z_i,a/r_I) - f_{1,f}(z_i,0).
\label{eq-deltaf}
\end{equation}

In this work we only focus on the fermionic term $4\pi\delta^{(1,f)}_F$ of Eq.~\eqref{eq-deltaf}, as the calculation of the gluonic term becomes much more intricate when using HYP-smeared links and it would need some extra care\footnote{For the calculation of the gluonic term with unsmeared links, we refer to \cite{Bali:2002wf}.}. Moreover, as we will see in Section \ref{sec-cont-limit}, our continuum extrapolation of $\alpha_{qq}$ is already accurate enough and allows to clearly distinguish the charm-loop effects on $\alpha_{qq}$ at high energies. This part of the calculation has been carried out using our computer package written in Mathematica. The fermionic contribution to the static force has been calculated for a sequence of finite lattice sizes ($L=[16,64]$) and then extrapolated to infinite volume.
\subsection{One-loop cutoff effects}
In this section we summarize our numerical results concerning the extraction of the leading fermionic lattice artifacts. The calculation has been realized for unsmeared and HYP2-smeared links, in order to see what are the main differences between these two cases. 

\begin{figure}[tp]
   \centering
   \subfigure
             {\includegraphics[width=0.48\textwidth,clip]{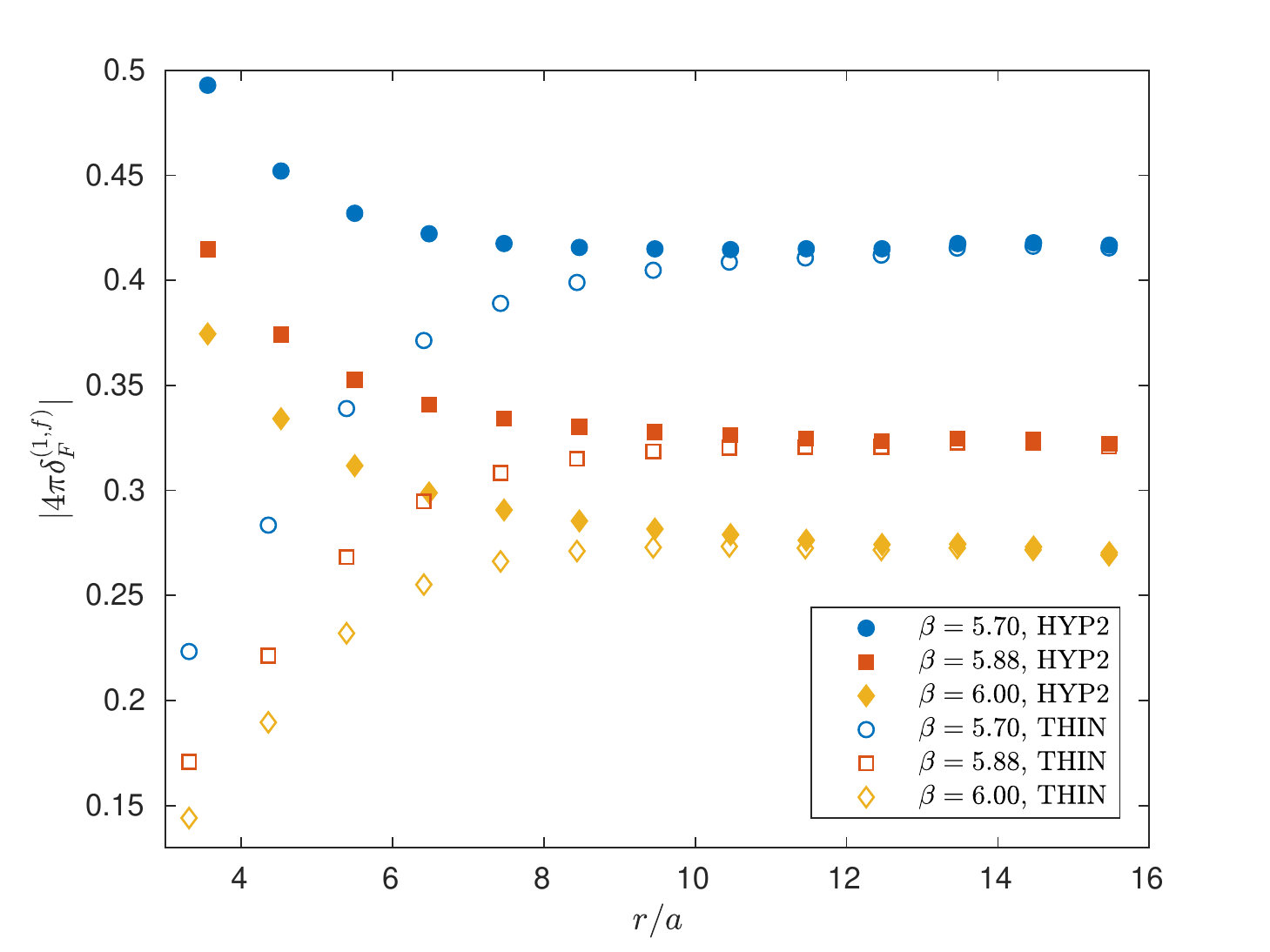}}\hfill
   \subfigure
             {\includegraphics[width=0.48\textwidth,clip]{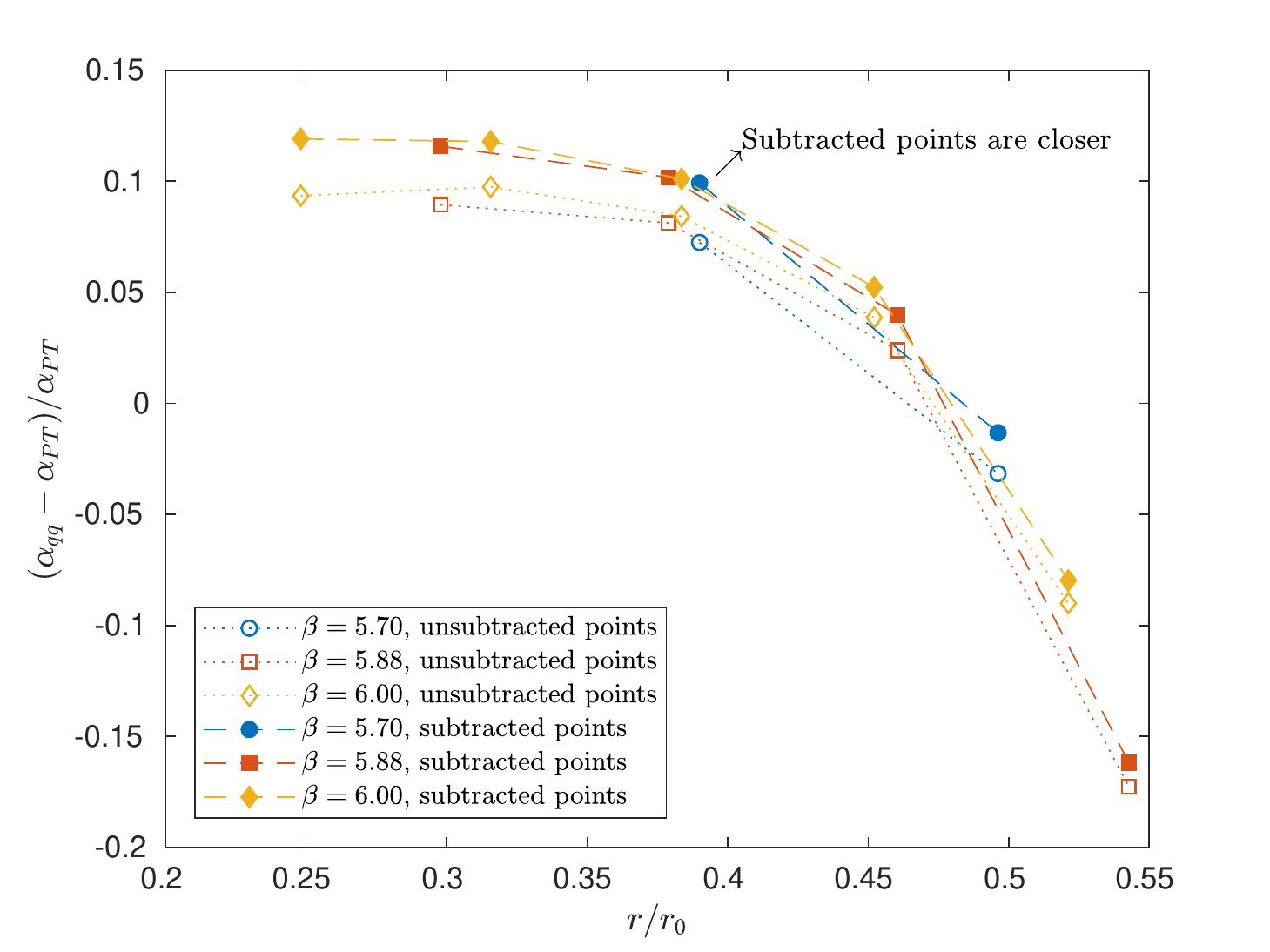}}
   \caption{The left panel shows the calculation of $4\pi\delta_F^{(1,f)}$ for unsmeared (empty markers) and HYP2-smeared links (full markers) choosing the same lattice discretization as our dynamical ensembles. In the right panel we subtract the leading lattice artifacts from our non-perturbative data of $\alpha_{qq}$ (full markers) and we compare to the unsubtracted ones (empty markers). $\alpha_{PT}$ is a perturbative 4-loop calculation of $\alpha_{qq}$ in QCD ($N_f=0$), using the $\Lambda$ parameter from \cite{Capitani:1998mq}.}
\label{fig:lat-art}
\end{figure}

On the left hand side of Figure \ref{fig:lat-art} we show the calculation of $4\pi\delta_F^{(1,f)}$ using the lattice parameters of our dynamical ensembles. We see clearly that at large distances in lattice units unsmeared and HYP2-smeared links produce similar effects, while at small distances using HYP2-smeared links gives rise to larger lattice artifacts. Moreover, as we expect, these unwanted effects become smaller when $\beta\rightarrow\infty$. To obtain the relative size of the lattice artifacts, $4\pi\delta_F^{(1,f)}$ must be multiplied by $g^2_{\overline{MS}}$ (see Eq. (\ref{eq-rel-lat-art})) and this means, at our lattice spacings, that we observe around $5\%$ effects.

The second step is to subtract the leading lattice artifacts from our non-perturbative data of $\alpha_{qq}$. The result of this procedure is depicted in Figure \ref{fig:lat-art} (r.h.s), where also a comparison with the unsubtracted data is shown. The most evident effect of this subtraction is that data corresponding to similar physical distances are closer to each other compared to the unsubtracted ones, as we can see clearly from the measurements of $\alpha_{qq}$ at around $0.38\mbox{ }r/r_0$. This is encouraging because it is a signal that the size of the lattice artifacts has been reduced considerably.
\section{Continuum limit}\label{sec-cont-limit}
The last part of this work is aimed at extracting the continuum limit of $\alpha_{qq}$ in QCD ($N_f=2$) at $M=M_c$ and QCD ($N_f=0$), in order to evaluate how large the charm loop effects are on this observable.
\subsection{Strategy}
Studying the \textit{step scaling function} $\sigma$ \cite{Luscher:1991wu} can provide a powerful tool to reach this purpose. If $f$ is a \textit{fixed scale factor}, $\sigma(f,u)$ measures how much the coupling changes when the distance scale changes by a factor $f$
\begin{equation}
\sigma(f,u) = g^2_{qq}(f\times r)|_{g^2_{qq}(r)=u}.
\end{equation} 
Then, from the definition of the $\beta_{qq}$-function
\begin{equation}
\beta_{qq} = -r\frac{\partial g_{qq}}{\partial r},
\label{eq-betaqq}
\end{equation}
one arrives at the exact relation
\begin{equation}
\log\left(f\right) = - \int_{\sqrt{u}}^{\sqrt{\sigma(f,u)}}\,\frac{dx}{\beta_{qq}(x)} .
\label{eq-exact-relation-sigma}
\end{equation}
Eq. (\ref{eq-exact-relation-sigma}) is only true in the continuum, but following \cite{DallaBrida:2016kgh} we parametrize $\beta_{qq}$ and the cutoff effects in such a way that the continuum limit of $\alpha_{qq}$ can be easily extracted from our lattice simulations.

Let us introduce the following parametrization of $\beta_{qq}$, which is not motivated by perturbation theory, but it allows to parametrize our data really well:
\begin{equation}
\beta_{qq} = -\frac{g^3_{qq}}{P(g^2_{qq})},\quad P(g_{qq}) = p_0 + p_1g^2_{qq} + p_2g^4_{qq} +\cdots.
\label{eq-param-beta}
\end{equation}
This choice permits to rewrite Eq. (\ref{eq-exact-relation-sigma}) as
\begin{equation}
\log\left(f\right) = -\frac{p_0}{2}\left[\frac{1}{\sigma(f,u)} - \frac{1}{u} \right] + \frac{p_1}{2}\log\left[\frac{\sigma(f,u)}{u}\right] + \sum_{n=1}^{n_{max}}\frac{p_{n+1}}{2n}\left[\sigma^n(f,u) - u^n\right],
\label{eq-param-sigma}
\end{equation}
where $2\times(n_{max} + 1)$ is the degree of the polynomial $P(g_{qq})$.

However, on a lattice one can only measure an approximation $\Sigma(f,u,a/r_0)$ of the step scaling function $\sigma(f,u)$ such that
\begin{equation}
\lim_{a\rightarrow 0}\Sigma\left(f,u,a/r_0\right) = \sigma(f,u).
\end{equation}
This means that if want to extract the coefficients $p_0, p_1,\dots, p_{n_{max}+1}$ of our parametrization of $\beta_{qq}$ from lattice simulations, instead of Eq. (\ref{eq-param-sigma}) we have to use
\begin{equation}
\log\left(f\right) + h(f,u,a/r_0) = -\frac{p_0}{2}\left[\frac{1}{\Sigma(f,u,a/r_0)} - \frac{1}{u} \right] + \frac{p_1}{2}\log\left[\frac{\Sigma(f,u,a/r_0)}{u}\right] + \sum_{n=1}^{n_{max}}\frac{p_{n+1}}{2n}\left[\Sigma^n(f,u,a/r_0) - u^n\right],
\label{eq-bestfit1}
\end{equation}
where $h(f,u,a/r_0)$ is a particular function that depends, other than $f$ and $u$, on the lattice spacing $a$. Since we expect cutoff effects proportional to $a^2/r_0^2$, we choose to parametrize $h(f,u,a/r_0)$ as
\begin{equation}
h(f,u,a/r_0) = \rho(f,u)\times\frac{a^2}{r_0^2},\quad \rho(f,u) = \sum_{i=0}^{n_{\rho}-1}\rho_i(f)u^i.
\label{eq-bestfit2}
\end{equation}

Thus, we can estimate the coefficients $p_i$ of $\beta_{qq}$ performing a global best-fit to our data at different lattice spacings through the Eqs.~\eqref{eq-bestfit1} and \eqref{eq-bestfit2}. The continuum extrapolation of $\alpha_{qq}$ can be realized solving the ODE which follows from the definition of the $\beta_{qq}$-function (Eq. (\ref{eq-betaqq})). Since this requires the choice of an initial condition (or the value of the $\Lambda$ parameter for both theories), we need to know the value of $\alpha_{qq}$ at a reference distance $r_{ref}$ in the continuum limit. This can be achieved by using an interpolation function for the static force \cite{Necco:2001xg}
\begin{equation}
F(r_{ref}) = f_1 + f_2r_{ref}^{-2}
\end{equation} 
between the two neighboring points. Setting $r_{ref}=0.75r_0$ and taking the continuum limit of the interpolations realized at different lattice spacings we obtain
\begin{itemize}
\item $N_f=0$: $\alpha_{qq}(0.75r_0)=0.7947(28)$, $\quad\chi^2/N_{dof}=0.13$ (constant fit); 
\item $N_f=2$: $\alpha_{qq}(0.75r_0)=0.8076(20)$, $\quad\chi^2/N_{dof}=0.27$ (constant fit).
\end{itemize}

\subsection{Results of the best-fits}
Before showing our final results, we begin this section with a few remarks. Using improved distances $r_I$, it is not possible to keep the factor $f$ exactly constant. However, we have seen that choosing $f\in [2.03,2.10]$ and $f\in [1.94,2.10]$ for our quenched and dynamical ensembles respectively produces acceptable best-fits for a large number of parametrizations of $\beta_{qq}$ and $\rho(u)$, see Eqs.~\eqref{eq-param-beta}, \eqref{eq-bestfit2}. The widths of these ranges depend in some manner on the lattice spacings and on the explored distances, therefore some attempts were needed before arriving at the ranges above-mentioned.

We tried different types of (correlated) best-fits, varying the number of parameters both in the parametrization of the $\beta_{qq}$-function and in the function $\rho(u)$ that parametrizes the cutoff effects.
We only show the parametrizations that provide an acceptable chi-squared using the minimum number of parameters, underlining that we obtain compatible results for different best-fits. 

\begin{table}[thb]
  \small
  \centering
  \begin{tabular}{ccccccccc}\toprule
  Theory  & $p_0$ & $p_1$ & $p_2$ & $p_3\times 10^2$ & $p_4\times 10^3$ & $p_5\times 10^5$ & $\rho_0$ & $\frac{\chi^2}{N_{dof}}$\\\midrule
  $N_f=0$ & $16.07(78)$ & $-3.33(44)$ & $0.610(86)$ & $-4.04(71)$ & $1.21(25)$ & $-1.32(31)$ & $0.96(43)$ & $\frac{21.31}{20}$\\\midrule
  $N_f=2$  & $14.64(50)$ & $-1.99(22)$ & $0.308(32)$ & $-1.32(18)$ & $0.192(33)$ &  & $0.85(26)$ & $\frac{24.67}{24}$\\\midrule
  $N_f=2$ \\ (subtracted) & $15.84(52)$ & $-2.25(23)$ & $0.329(33)$ & $-1.38(18)$ & $0.198(33)$ &  & $0.22(25)$ & $\frac{18.17}{24}$\\\bottomrule
  \end{tabular}
  \caption{Results of the continuum extrapolation in $N_f=0$ and $N_f=2$ theories. In the second row we show the results obtained with the original non-perturbative data, while in the third row the ones obtained subtracting the leading fermionic lattice artifacts.}
\label{tab-bestfit-pars}
\end{table}

Table \ref{tab-bestfit-pars} summarizes the results of our continuum extrapolations. The numbers listed in the table have been produced using $6$ parameters for $\beta_{qq}$ and $1$ for $\rho(u)$ in quenched QCD (thus $6+1$ parameters on the whole), whilst $5+1$ parameters have been used in QCD ($N_f=2$). For equal number of parameters, subtracting the leading fermionic lattice artifacts for the $2$ flavor theory produces a bit smaller relative errors and a better chi-squared. Moreover, we can see that the coefficient $\rho_0$, introduced to parametrize the cutoff effects, is compatible with zero when the one-loop fermionic lattice artifacts are subtracted. Expanding \eqref{eq-param-beta} in powers of $g_{qq}$, we can rewrite $\beta_{qq}$ in a way similar to the one motivated by perturbation theory. In particular, the first two coefficients are given by
\begin{equation}
\beta_{qq} = -g^{3}_{qq}\left(\frac{1}{p_0} - \frac{p_1}{p_0^2}g^2_{qq} \right) + \mathcal{O}(g^7_{qq}) \equiv -g^{3}_{qq}\left(b_0 + b_1g^2_{qq} \right) + \mathcal{O}(g^7_{qq}).
\end{equation}
This allows us to identify $b_0\equiv 1/p_0$ and $b_1\equiv-p_1/p_0^2$. From our continuum extrapolation of the quenched theory we obtain $b_0=0.062(3)$ and $b_1=0.013(3)$. These estimates deviate a bit from the perturbative results for $N_f=0$ ($b_0=\frac{11}{(4\pi)^2}\approx 0.070$ and $b_1=\frac{102}{(4\pi)^4}\approx 0.004$), but this is something that we could expect because the parameters $p_i$ have been estimated in a range which is far away from the domain of validity of perturbation theory ($\alpha_{qq}<0.20$ in case of a 4-loop calculation).
\begin{figure}[tp]
   \centering
   \subfigure
             {\includegraphics[width=0.48\textwidth,clip]{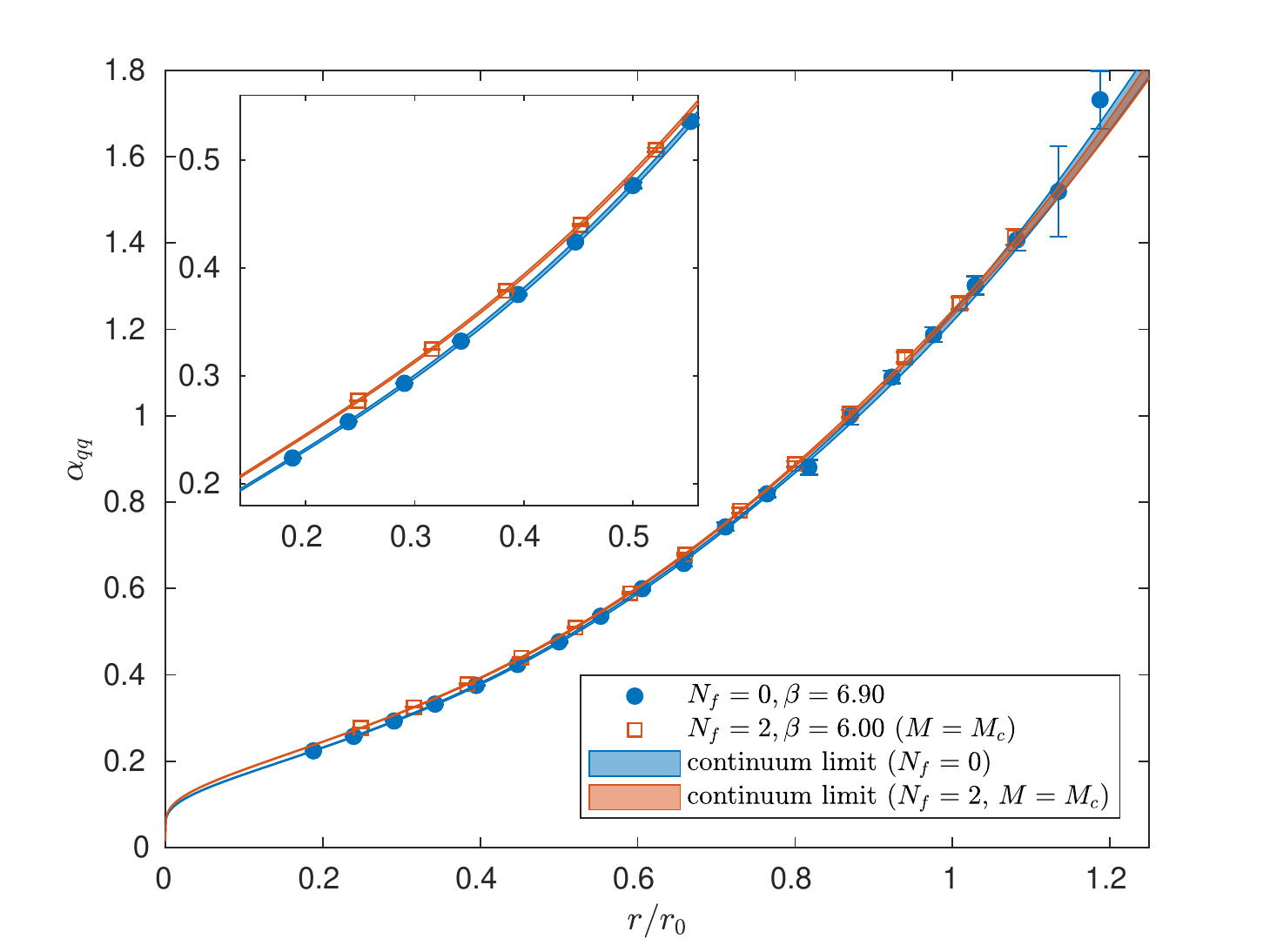}}\hfill
   \subfigure
             {\includegraphics[width=0.48\textwidth,clip]{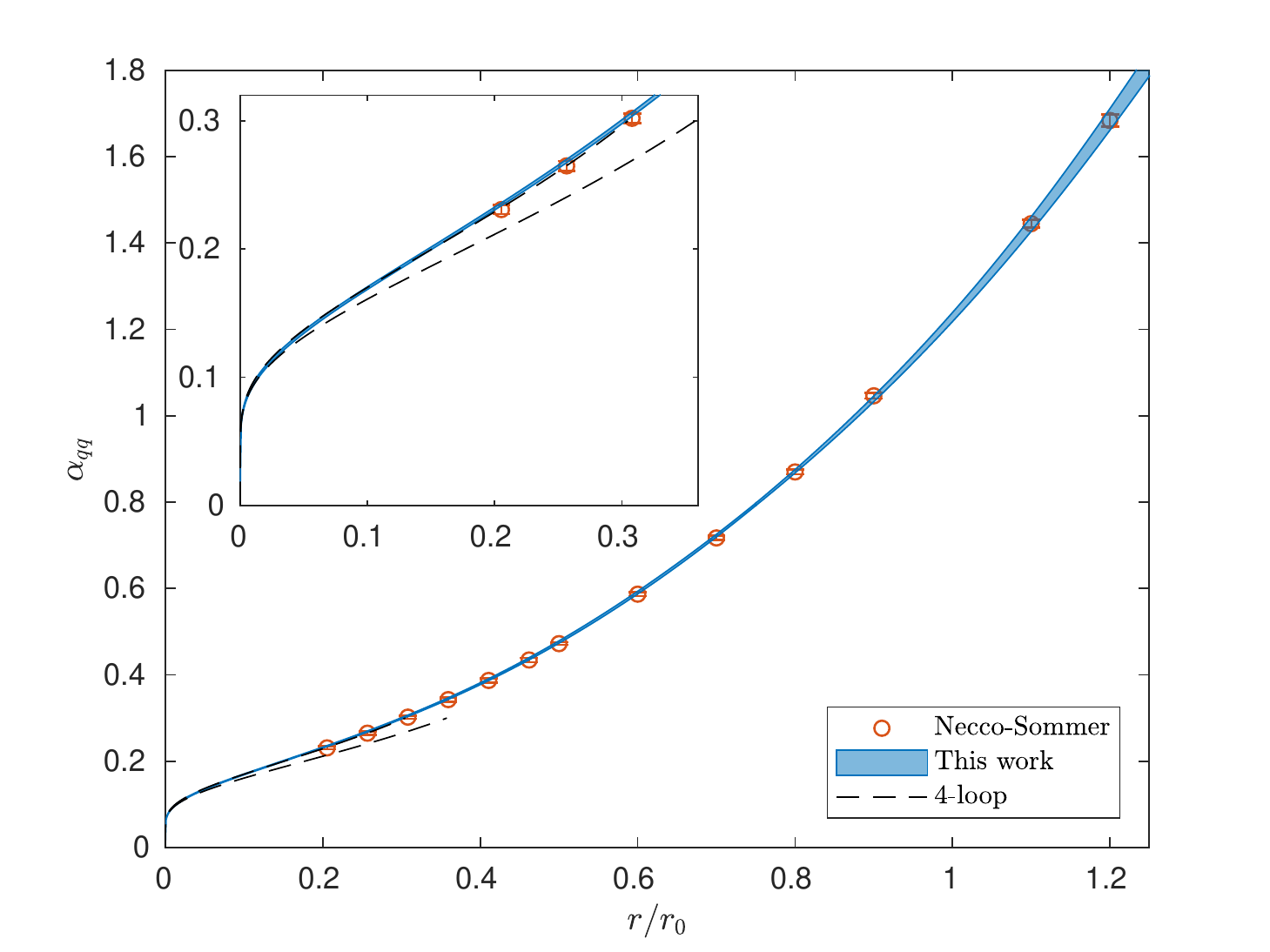}}
   \caption{In the left panel a comparison of $\alpha_{qq}$ in QCD ($N_f=2$) at $M=M_c$ and quenched QCD is shown. The blue circles and the red squares denote the measurements of $\alpha_{qq}$ produced with our finest lattices in $N_f=0$ and $N_f=2$ theories respectively. The blue and red bands stand for the continuum extrapolations obtained using three different lattice spacings for both theories. The widths of the bands originate from the errors on the data and their correlation is taken into account. The right panel shows instead a comparison between our continuum extrapolation of $\alpha_{qq}$ in quenched QCD (the blue band) and the one obtained in a previous work \citep{Necco:2001xg} (the red circles) using a different strategy. The dashed black lines are the predictions in perturbation theory up to four loops. The spread of the lines comes from the uncertainty in the $\Lambda$ parameter. }
\label{fig:alphaqq}
\end{figure}
Once we know the parameters of the $\beta_{qq}$-function, the continuum limit of $\alpha_{qq}$ can be easily extracted and our final results are depicted in Figure \ref{fig:alphaqq}. On the left hand side a comparison of $\alpha_{qq}$ in $N_f=0$ and $N_f=2$ (subtracting the leading fermionic lattice artifacts) theories is shown, where we use $r/r_0(M_c)$ on the x-axis for the dynamical points. We see that the continuum limits are accurate enough to distinguish the dynamical charm effects on $\alpha_{qq}$ at distances $r/r_0\lesssim0.5$. In the right panel we compare our continuum extrapolation of $\alpha_{qq}$ in quenched QCD to the one obtained in Ref. \cite{Necco:2001xg} and the predictions of perturbation theory up to four loops. We observe a really good agreement with \cite{Necco:2001xg} and with perturbation theory at $\alpha_{qq}\lesssim0.20$.
\section{Conclusions and Outlook}
In this work we have focused on evaluating the one-loop cutoff effects of a dynamical charm quark on $\alpha_{qq}$ using HYP2-smeared links. We see that these effects are small, but HYP2-smeared links produce bigger lattice artifacts compared to unsmeared links at small distances in lattice units.

We have also tried to extract the continuum limit of $\alpha_{qq}$ studying the step scaling function $\sigma$. This strategy allowed us to evaluate the dynamical charm effects on $\alpha_{qq}$ and for quenched QCD we find a really good agreement with perturbation theory at high energies and Ref. \cite{Necco:2001xg} at low energies. Our continuum extrapolations also indicate that the dynamical charm effects on $\alpha_{qq}$ are significant at distances $r/r_0\lesssim0.5$.

In future we plan to extend these measurements to other values of quark masses to study the mass-dependence of the strong coupling $\alpha_{qq}$. 

\section*{Acknowledgments}
We gratefully acknowledge the Gauss Centre for Supercomputing (GCS) for providing computing time on the supercomputers JURECA and JUQUEEN at J\"ulich Supercomputing Centre (JSC). S.C. acknowledges support from the European Union's Horizon 2020 research and innovation programme under the Marie Sk\l{}odowska-Curie grant agreement No. 642069.

\bibliography{lattice2017}

\end{document}